\documentclass[apj]{emulateapj}
\usepackage{graphicx,amssymb,amsmath}
\usepackage{booktabs}
\usepackage{ulem,color}
\usepackage{natbib}
\usepackage{pifont}
\usepackage[flushleft]{threeparttable}

\def\apjl{ApJL }
\def\aj{AJ }
\def\apj{ApJ }

\def\aap{A\&A }
\def\nat{Nature }
\def\mnras{MNRAS }
\def\prd{Phys. Rev. D. }

\def\cM{{\cal M}_*}

\begin{document}

\author{Tsvi Piran$^1$,  Gilad Svirski$^1$, Julian Krolik$^2$, Roseanne M. Cheng$^2$  and Hotaka Shiokawa$^2$}
\affil{1. Racah Institute of Physics, The Hebrew University of Jerusalem, Jerusalem 91904, Israel
\\
2. Physics and Astronomy Department, Johns Hopkins University, Baltimore, MD 21218, USA}
\title{``Circularization" vs.   Accretion -- What Powers Tidal Disruption Events? }

\begin{abstract}
A tidal disruption event (TDE) takes place when a star passes near enough to a massive black hole to be disrupted.
About half the star's matter is given elliptical trajectories with large apocenter distances, the other half is unbound. To ``circularize", i.e., to form an accretion flow, the bound matter must lose a significant amount of energy, with the
actual amount depending on the characteristic scale of the flow measured in units of the black hole's gravitational radius ($\sim 10^{51} (R/1000R_g)^{-1}$~erg).  Recent numerical simulations \citep{Shiokawa+2015} have revealed that the circularization scale is close to the scale of the most-bound initial orbits,  $\sim 10^3 M_{BH,6.5}^{-2/3} R_g \sim 10^{15}  M_{BH,6.5}^{1/3}$~cm from the black hole, and the corresponding circularization energy dissipation rate is $\sim 10^{44} M_{BH,6.5}^{-1/6}$~erg/s.  We suggest that the energy liberated during  circularization, rather then energy liberated by accretion onto the black hole, powers the observed optical TDE candidates.    The  observed rise times, luminosities, temperatures, emission radii, and line widths seen in these TDEs  \citep[e.g.][]{Arcavi+2014} are all more readily explained in terms of heating associated with circularization than in terms of accretion.
\end{abstract}

\section{Introduction}\label{sec:intro}

Occasionally, a star passes close enough to a supermassive black hole (SMBH) to be tidally disrupted. The  frequency of such occasions is estimated to be $\sim 10^{-5}$ to $10^{-4}$ events per galaxy per year \citep[e.g.][]{MagorrianTremaine99,Donley+2002,Wang.Merritt.2004,Kesden2012,Stone2014,vanVelzen2014}. 
Although the first tidal disruption event (TDE) candidates were discovered in the X-rays or UV \citep[e.g.][]{Komossa+2004,Gezari+2009,Bloom+2011}, a growing number of TDE candidates have been recently discovered in the optical \citep[e.g.][]{Gezari.et.al.2012,Chornock+2014,Holoien+2014,vanVelzen2014,Arcavi+2014}.  However, the recent optical TDE observations are difficult to reconcile with theoretical expectations  \citep[e.g.][]{Ulmer99,Strubbe.Quataert.2009,Lodato2011},  in which the optical signal is due to accretion onto the black hole. 
 
The classical description of a  TDE was outlined by \cite{Rees.1988}.  In this picture, a star of mass $M_* $ and radius $R_*$ approaches a supermassive black hole with mass  $M_{\rm{BH}}$.  The star is disrupted when its parabolic trajectory brings it to a pericenter distance $R_p$ smaller than the tidal radius, $R_T$.   The resulting stellar debris has a rather narrow distribution of specific angular momentum (all of it close to the specific angular momentum of the star), but the distribution of mass with respect to specific binding energy $dM/d\epsilon$ is roughly flat from $\epsilon \simeq -GM_{\rm{BH}}R_*/R_T^2$ to $\epsilon \simeq +GM_{\rm{BH}}R_*/R_T^2$.     Because the semi-major axis of the most-bound matter is
\begin{equation}
a_{\rm min} \simeq (1/2) R_T^2/R_* \sim  (M_{BH}/M_*)^{1/3} R_T \gg R_T \,
\end{equation}
the orbits of these ``tidal streams" are highly eccentric.    A uniform distribution of mass per binding energy implies a mass return rate to the stellar pericenter $\dot{M}_{\rm{fallback}}\propto ({t}/{t_0})^{-5/3}$, where  $t_0$  is the orbital  time of the most bound material  \citep{Phinney.1989}.
In this model it is further assumed that upon return to pericenter, general relativistic apsidal precession causes streams returning at different times to shock against each other and dissipate sufficient orbital energy to compress these very extended, highly elliptical orbits into approximately circular orbits with radii $\sim 2R_p$.    The inflow time through the accretion disk that then forms is estimated to be $\ll t_{0}$, so that the accretion rate onto the black hole tracks closely the mass return rate of the tidal streams.
If so, the bolometric light curve should peak at $\sim t_0 \sim 2 \times 10^6 M_{\rm BH,6.5}^{1/2}$~s (where $M_{\rm BH,6.5} \equiv M_{\rm BH}/10^{6.5} M_\odot$, we choose this mass as fiducial because it best matches the black hole masses estimated from stellar bulge properties for the events observed by \citealt{Arcavi+2014}) after the star is destroyed, reaching a maximum luminosity $\sim 2\times 10^{46}  M_{BH,6.5}^{-3/2}$~erg/s and then decay $\propto (t/t_0)^{-5/3}$.   
This temporal decay became the hallmark for observational identification of TDEs, and 
indeed several TDE candidates have been reported as having such a light curve.
Within this model, the effective temperature of the peak would be $\sim 4 \times 10^7 M_{\rm BH,6.5}^{-7/2}$~K if light from the inner rings of the accretion disk reaches distant observers unimpeded; similarly, its effective radius would be $\sim 10R_g \sim 5 \times 10^{12}M_{\rm BH,6.5}$~cm.    Because the outer edge of the disk lies at a radius $\sim R_T$, even the narrowest emission lines coming from the accretion flow proper would have extremely large line widths, $\sim 0.2 M_{\rm BH,6.5}^{1/3}$.

This simple model faces now serious problems when confronted with observations of optical TDE candidates \citep{Gezari.et.al.2012,Chornock+2014,Holoien+2014,vanVelzen2014,Arcavi+2014}. Optical light curves of these events show (in rough terms) the expected $t^{-5/3}$ decline, and the rise time agrees with the return time expected from a TDE due to a $\sim 5 \times10^6 M_\odot$ SMBH. 
However, the observed temperature and bolometric luminosity (2--$3 \times 10^4$~K, $\sim 10^{43}$--$10^{44}$~erg/s, taking at face value the published bolometric correction) are significantly lower
 than predicted, while  the inferred black body emission radius is much larger ($\sim 10^{15}$~cm, i.e., $\sim 2 \times 10^3 R_g M_{\rm BH,6.5}^{-1}$).  In addition, line broadening, which reflects orbital motion, is, at $\sim 10,000$~km/s, much smaller than would be expected for an accretion disk on the scale of $R_T$.    These results bear little resemblance to the classical model's predictions  \citep[as noted also by][and others]{Guillochon.Manukian.Ramirez-Ruiz2014, Chornock+2014,Arcavi+2014}.
 
The assumptions behind the classical TDE picture have been criticized by many authors.  Some simulations find that the heating associated with inter-stream shocks unbinds a significant part of the gas \citep[e.g.][]{Ayal+2000}.    If, as is expected for the parameters of many TDE events, the peak accretion rate is super-Eddington \citep{Loeb.Ulmer.1997}, the accretion luminosity itself might power a wind that expels gas \citep{Ohsuga+2005,Dotan.Shaviv2011}.   Such outflows may significantly affect the observed light curve \citep[e.g.][]{Strubbe.Quataert.2009, Lodato2011}.  \cite{Lodato.et.al.2009}  and \cite{Guillochon.Ramirez-Ruiz.2013} showed that, depending on the detailed structure of the star, ${dM}/{d\epsilon}$ is not necessarily constant, leading to further deviations from the simple picture.  Finally, even if the bolometric light curve does follow a $t^{-5/3}$ decay, the optical light curve should not reflect the bolometric one if the spectrum peaks, as it often should, in the EUV \citep{Lodato2011}.

However, there is another aspect of the classical TDE picture that has thus far escaped much critical review: the
assumption that the gas circularizes immediately upon returning to the vicinity of the black hole, and that it does so on
the scale of the tidal radius.    To circularize the returning matter at a radius $R < a_{\rm min}$, the gas must lose an energy
$\sim GM_{\rm BH}/R$ per unit mass.    Dissipation (e.g., in shocks) can help, but only if heating quickly leads to true energy
loss by radiation that escapes the matter.   Moreover, in order to dissipate this much energy, the shocks must have speeds comparable to the orbital speed at radius $R$.    In the earliest work, it was supposed that these shocks would take place near the pericenter (i.e. at a radius from the black hole $\simeq R_p$ and at approximately the azimuthal angle where the star reaches $R_p$), and would be caused by strong relativistic apsidal precession \citep{Rees.1988}.    Reconsideration by
\cite{Kochanek1994} indicated that any shock near pericenter would be much weaker, with the dominant stream convergence due to the small angular spread of the tidal streams rather than to apsidal precession.
In a pseudo-Newtonian SPH simulation covering the beginning of mass return, \cite{Rosswog.et.al.2009} also found shocks between the tidal streams near apocenter, but because the available kinetic energy is least at apocenter, these shocks dissipate relatively weakly.
Despite these qualms, it has still been generally assumed that somehow the tidal debris would quickly ``circularize", i.e., lose enough of its orbital energy through hydrodynamical processes that it can join an accretion disk with
a radial extent $\sim R_T$.

Recently  \cite{Shiokawa+2015} took up this question using detailed numerical simulations.
(Because of numerical reasons, these simulations were carried out for a tidal disruption of white dwarf by an intermediate mass black hole, but the simulations can be scaled to 
a regular star tidally disrupted by a massive black hole.)
They found that the circularization process is slower than previously thought (lasting $\simeq (5$--$10)t_0$), produces a flow that is only roughly circular ($\langle e \rangle \simeq 0.4$), and leaves most of the debris at radii nearer $a_{\rm min}$ than $R_T$ because the principal shocks are located at that scale.   By quantitatively defining ``circularization" in terms of progress toward reducing the eccentricity of the fluid orbits and making the flow axisymmetric, they demonstrated that circularization may remain incomplete even at the end of the event (hence the quotes around the term in the title of this paper).    Even more importantly, they pointed out that if the characteristic radius of the gas is significantly greater than $R_T$, inflow could take considerably longer than $\sim t_0$, decoupling the time-dependence of the light output from the time-dependence of the mass-return rate.

In this paper, we build upon their analysis of where the shocks in this flow are located and the magnitude of the heating rate associated with them to point out that the very fact they occur at rather larger radius than previously expected makes them a strong candidate for the origin of the light seen in optical TDE events \citep{Arcavi+2014}.    As we will show, the peak luminosity, the color temperature (and therefore the size of the emitting region), and the line widths are all reproduced well by this model.

We use the term ``circularization\footnote{Note that  \cite{Lodato12} and \cite{Bogdanovic+2014} have used the term 
circularization in the context of TDE lightcurves, but neither one was referring to the radiation of heat generated in shocks at large radius.}" as a label for our outer shocks-powered TDE model in order to distinguish it from the commonly assumed accretion-powered model.   However, we emphasize that in fact the gas orbits remain highly elliptical even after passing through these shocks because their binding energy remains much smaller than that corresponding to a circular orbit with their angular momentum.    In addition, the gas is partially pressure-supported.

We outline the results of the numerical simulations  and the apocentric circularization model stemming from them in
\S~\ref{sec:model}. We summarize the observations \citep{Gezari.et.al.2012,Chornock+2014,Holoien+2014,vanVelzen2014,Arcavi+2014} and  compare them to the predictions of this model in \S~\ref{sec:obs}.  We conclude and summarize the implications of these findings in \S \ref{sec:conc}. 
  
\section{The circularization model} 
\label{sec:model} 

We begin by briefly summarizing a few basic properties of tidal disruption events. 
For a main sequence star, the radius $R_* = R_{\odot} \cM^{1-\xi}$, where $\cM \equiv M_*/M_{\odot}$ and $\xi \approx 0.2$ for $0.1 < \cM< 1$, but increases to $\approx 0.4$ for $1 < \cM < 10$ \citep{Kippenhahn1994}.   To account for the shape of the star's internal density profile, we follow \cite{Phinney.1989}, defining $f$ as the ratio of the gravitational binding energy of the star to $GM_*^2/R_*$ and $k$ as the apsidal motion constant (determined by the star's radial density profile). 
The ratio $k/f= 0.02$ for radiative stars, but is 0.3 for convective stars \citep{Phinney.1989}.    These ratios are extreme values, as most stars are a mix of radiative and convective regions: in stars less massive than the Sun, an outer convective zone surrounds a radiative core; in more massive stars, the core is convective while the envelope is radiative \citep{Kippenhahn1994}.   The quantitative balance between the two kinds of regions is sensitive to heavy element abundances because higher  increases opacity per unit mass, enlarging convective zones.   This could be a significant effect because the stars most likely to be the victims of TDEs are predominantly drawn from the central regions of galaxies, so their heavy element abundances may, on average, be several times greater than Solar \citep{Sarzi+2005,Rojas+2014,Gonzalez+2014}.  We will use in the following the geometric mean of the two values, $k/f=0.08$, as our fiducial value.

From the size of the star and a gauge of its internal structure (the $k/f$ ratio), we can estimate the tidal radius
\begin{equation}
R_T \approx 6.7 \times 10^{12} \, \left(\frac{k/f}{0.08}\right)^{1/6}\cM^{2/3-\xi}M_{\rm{BH,6.5
}}^{1/3}\,\rm{cm} \ .
\end{equation}
In the estimates to follow, we will assume $R_p \simeq R_T$ for 
several reasons: the cross section for disruptions is $\propto R_p$ and furthermore if the stellar loss-cone isn't full, the rate of smaller $R_p$ encounters is suppressed \citep{FrankRees1976};  additionally 
the  debris energy distribution is unchanged for
$R_p < R_T$ \citep{Stone.Sari.Loeb2013,Guillochon.Ramirez-Ruiz.2013}.

The semi$-$major axis of the  most-bound material is:
\begin{equation}
a_{\rm min} \approx 3.2\times 10^{14}\,\left(\frac{k/f}{0.08}\right)^{1/3}\cM^{1/3-\xi}M_{\rm{BH,6.5}}^{2/3}\,\rm{cm}\ .
\label{Eq. amin}
\end{equation}
The corresponding return time of the most-bound material to the pericenter is:
\begin{equation}\label{eq:t0}
t_0 \approx 1.8\times 10^6\,\left(\frac{k/f}{0.08}\right)^{1/2}\cM^{(1-3\xi)/2}M_{\rm{BH,6.5}}^{1/2} \, \rm{s} ,
\end{equation}
and the maximal mass return rate is: 
\begin{equation}
\dot M_0 \approx  \frac{M_*}{3 t_0} \approx 3.6 \times 10^{26} \,\left(\frac{k/f}{0.08}\right)^{-1/2}\cM^{(1+3\xi)/2}M_{\rm{BH,6.5}}^{-1/2} \, {\rm gm \,s^{-1}} \ .
\end{equation}

\begin{figure}[b]
\includegraphics[width=9cm]{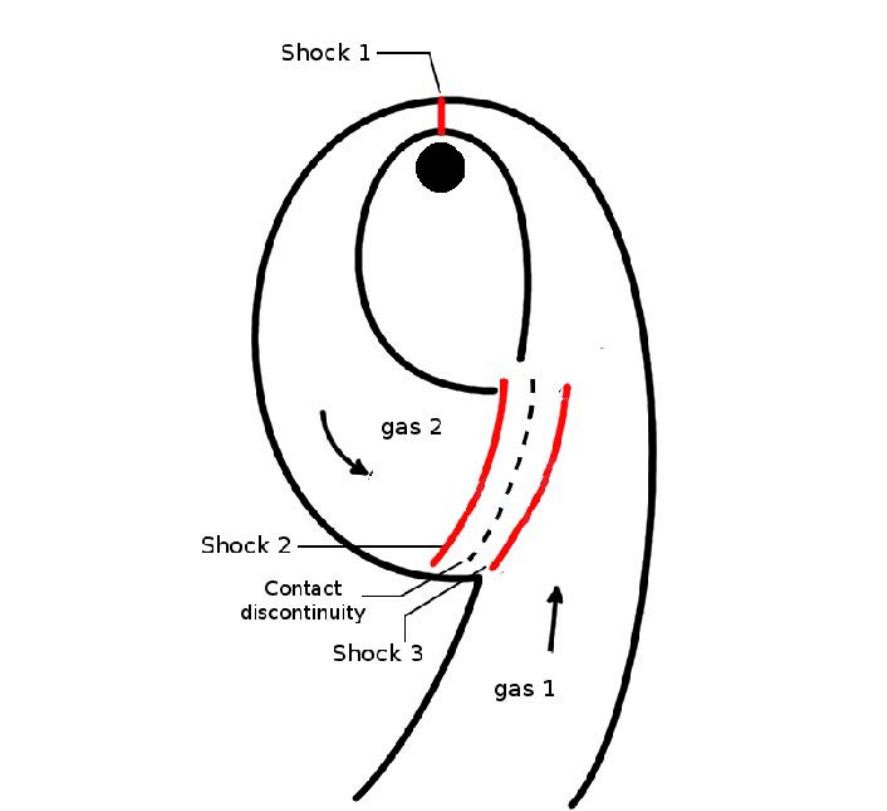}
\includegraphics[width=9cm]{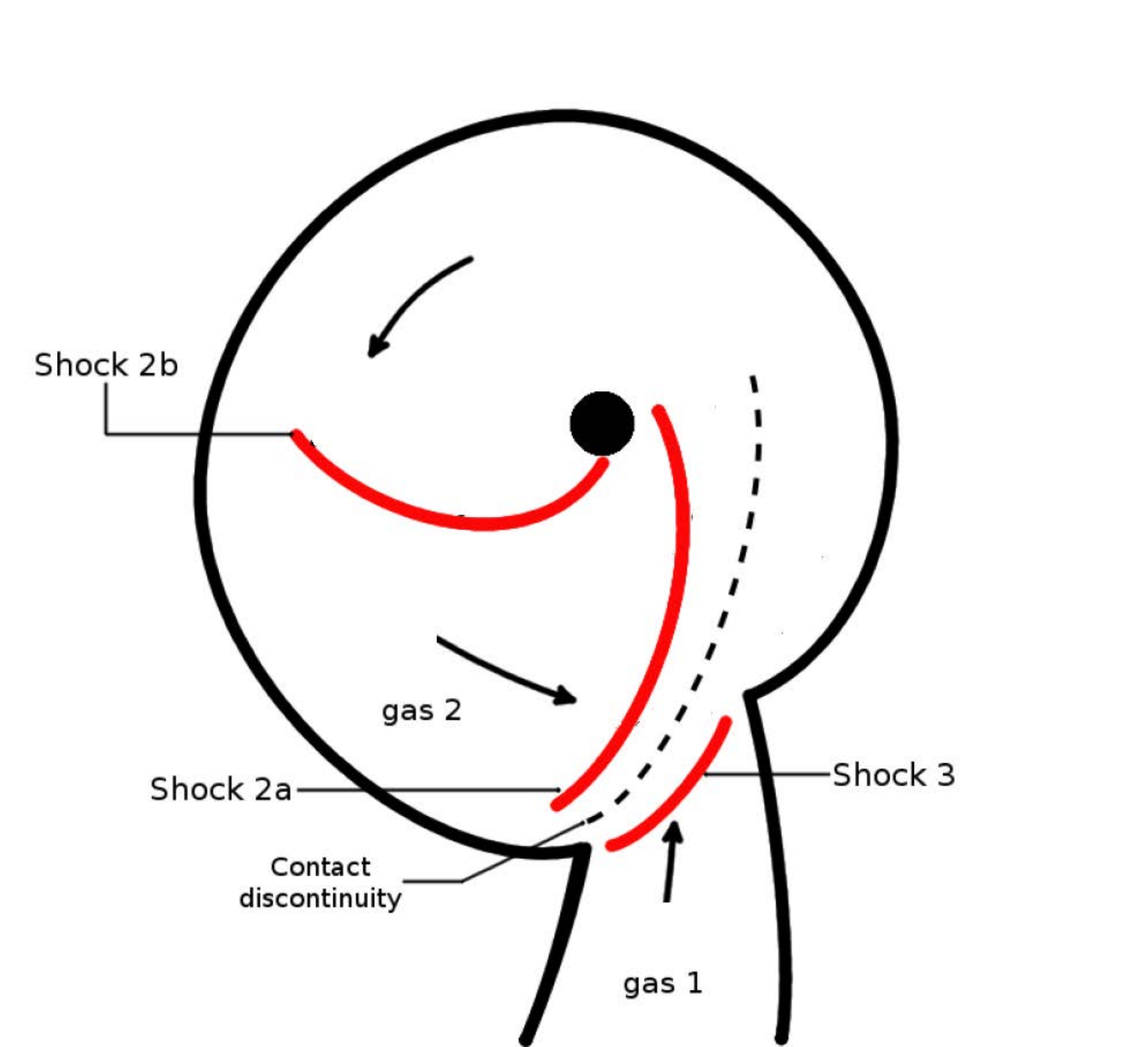}
\caption
{Schematic description of the shock system. The inner shock~1 takes place near the pericenter, while the outer shocks~2 and 3 arise nearer the apocenter. Shown are two different times, $t\simeq 2t_0$ (top) and $t \simeq 6t_0$ (bottom).   Note that
at earlier times shock~2 has only one branch (top), whereas it splits into two at later times (bottom).  Sample gas streamlines are
shown as black curves with arrows; shocks are shown in red.   The large black disk is the black hole; the thick
black curves without arrows are rough indications of the boundaries of the flow.   Gas-1 denotes returning matter, while gas-2 denotes matter that has gone around the black hole at least once. The contact discontinuity between gas-1 and gas-2 is indicated by a dashed black curve.  Separations are not drawn to scale in order to emphasize the sequence of events.
Adapted from \cite{Shiokawa+2015}.}
\label{Fig:shocks}
\end{figure}

Although \cite{Rees.1988} suggested that relativistic effects cause the apsidal angles for streams with different semi-major axes to reach the pericenter with such different directions that their mutual shocks have speeds of order the orbits' maximum orbital speed, quantitative study indicates this occurs only if $R_p/R_g$ is rather small or if the initial stellar orbit is bound \citep{Bonnerot+15,Hayasaki15}.   Similarly, Lense-Thirring precession also requires relatively small $R_p/R_g$ to be significant \citep{Guillochon.Ramirez-Ruiz2015,Hayasaki15}.  Instead, three separate shock systems (see Fig.~\ref{Fig:shocks}), none of them that strong, combine to transform the debris motion from highly-elliptical ballistic orbits to only moderately elliptical orbits significantly influenced by hydrodynamics \citep{Shiokawa+2015}.   Upon passing through the pericenter region, the geometrical convergence of streams in different orbital planes creates the ``nozzle shock" \citep{Evans.Kochanek.1989,Kochanek1994} dubbed ``shock~1" in \cite{Shiokawa+2015}.     Intrinsic misalignment of the orbital apsides between different stream orbits creates a forward/reverse shock system at radii $\sim a_{\rm min}$ where streams returning for the first time are intercepted by matter that has already passed through shock~1 at least once.    \cite{Shiokawa+2015} called the forward shock, the one in which fresh material is shocked, ``shock~3" and the reverse shock, the one acting on gas that had already returned, ``shock~2".    Over time, shock~2 divides into two arms, and  stream deflection, both by the outer shocks and by the increasing pressure in the pericenter region, leads to the disappearance of shock~1.   The mass return rate does roughly follow the classical $t^{-5/3}$ trend, but after 80\% of the bound debris had returned from apocenter, \cite{Shiokawa+2015} found that most of the mass is placed on only roughly circular orbits at radii 
$\sim a_{\rm min}$, $\simeq (5$--$8) R_T$ for their parameters, 
corresponding to 
$ a_{\rm min} \simeq 100 R_T$, when scaled to the tidal disruption by a SMBH, which is of interest here.  
Moreover, a time $\simeq (3$--$10) t_0$ is required for even this degree of ``circularization" to be achieved.  
  A fraction of the returning mass loses enough angular momentum by shock deflection that it is able to accrete within a few $t_0$, but the maximum accretion rate is only $\sim 0.1\times$ the classical expectation.
 
\cite{Shiokawa+2015} considered, for numerical reasons, 
a TDE of a white-dwarf by an IMBH. In that case $a_{\rm min}/R_p \sim 10$, and after redistribution of angular momentum most of the matter stays at a distance comparable to $a_{\rm min}$.  In a MS-SMBH this ratio is of order $\sim 100$, and clearly there is not enough angular momentum to keep the matter  at a circular 
orbit there. As such, the term ``circularization" might be somewhat confusing.   
However, because energy dissipation is so slow, the matter will still remain mostly at a distance comparable to $a_{\rm min}$.  It will be supported partially by thermal pressure, and it will settle into elliptical orbits.  
The near-apocenter location of the outer shocks still holds due to the geometry of elliptical orbits whose apsidal directions are slightly different and because the shocks are built upon collision with newly-arriving streams. The re-orientation of these newly-arriving streams at the location of the outer shocks weakens the inner shock and prevents further dissipation, and the streams remain highly eccentric in the absence of an additional mechanism for energy loss.

\begin{figure}
\includegraphics[width=9cm]{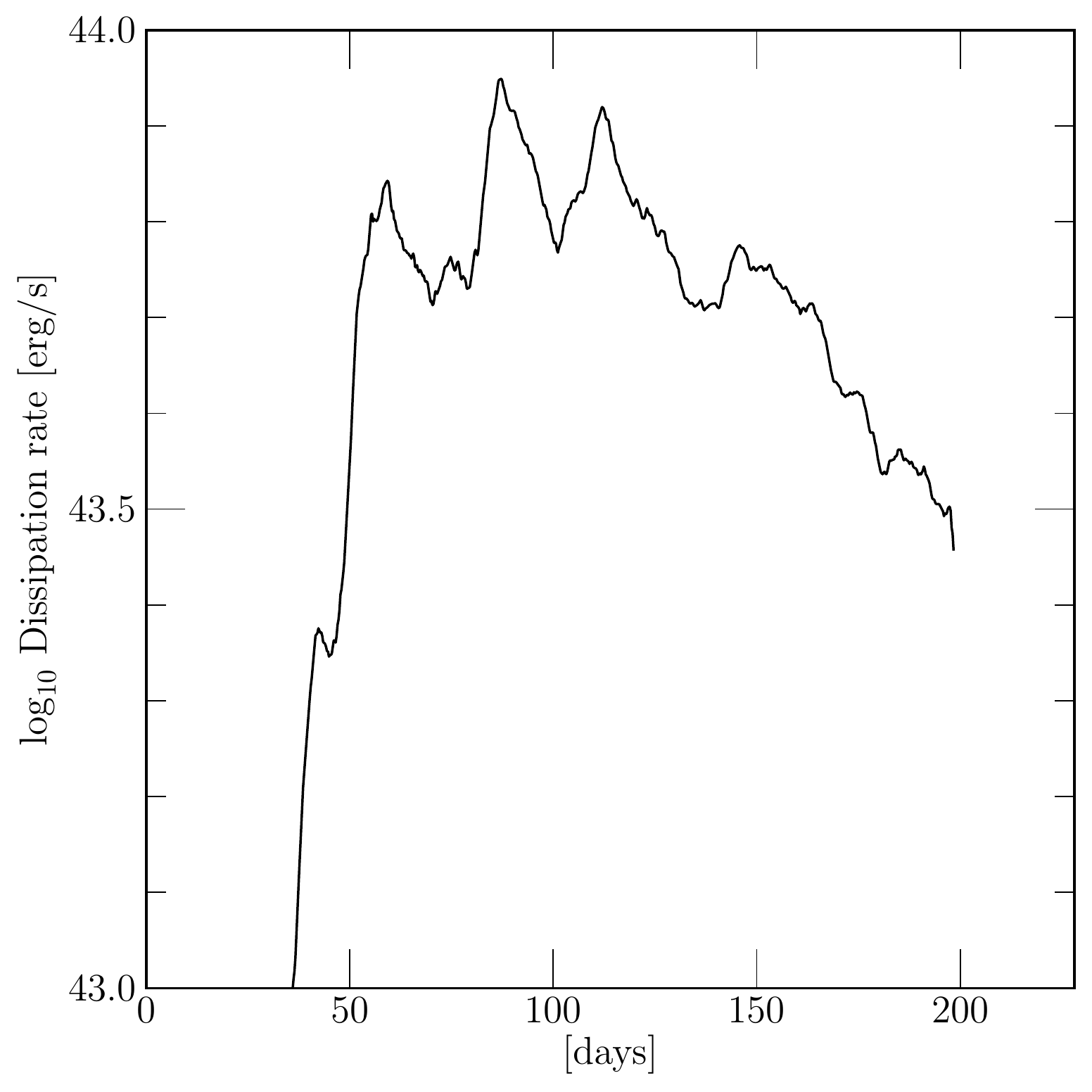}
\caption{Heating rate from  scaled to our fiducial main sequence parameters as described
in text.}
\label{fig:heatingrate}
\end{figure}

The outer shocks (shocks~2 and 3) dissipate energy at a rate comparable to the mass return-rate times the 
returning matter's orbital kinetic energy. Because the density of the matter approaching shock~3 from large radius is much greater (by as much as two orders of magnitude) than the density of matter that has already passed through shock~1 at least once, heating shocks~2 is rather greater than in shock~3 (see Fig.~13 \citet{Shiokawa+2015}).   To order of magnitude accuracy, the outer shock heating rate can be estimated by
\begin{equation}\label{eq:L_peak}
{\dot E}_{\rm peak}  \sim \frac{GM_{\rm{BH}} \dot M_0}{ a_{\rm{min}}}
\approx 5\times 10^{44} \left(\frac{k/f}{0.08}\right)^{-5/6}\cM^{{1}/{6}+{5\xi}/{2}}M_{\rm{BH,6.5}}^{-1/6}\,\rm{erg\,s^{-1}} ,
\end{equation}
The quality of this estimate can be confirmed by scaling the heating rate from the \citet{Shiokawa+2015} simulation,
which followed a $0.64 M_{\odot}$ white dwarf  (with $R_* = 0.0124 R_{\odot}$) disrupted by a $500M_{\odot}$ black
hole, to conditions appropriate to the fiducial values of this paper.   We do so assuming that
$\dot E_{\rm peak} \propto M_* (GM_{\rm BH}/a_{\rm min})/P_{\rm orb}(a_{\rm min})$.   The scaling can then be
most succinctly accomplished by writing $R_T = \Phi {\cal R} R_*$, with ${\cal R}= (M_{\rm BH}/M_*)^{1/3}$
and $\Phi = (k/f)^{1/6}$ for main sequence stars \citep{Phinney.1989}; our explicit calculation of the white dwarf disruption
indicates that $\Phi_{\rm WD}  \simeq 1.1$.   In this notation, the specific energy of the
most-bound debris $\epsilon = GM_{\rm BH} R_*/R_T^2 =  GM_{\rm BH}/(\Phi^2 {\cal R}^2 R_*)$, so that
$a_{\rm min} = GM_{\rm BH}/(2\epsilon) = \Phi^2 {\cal R}^2 R_*/2$ and $P_{\rm orb}(a_{\rm min}) =
(\sqrt{2}/\pi) \Phi^3 {\cal R}^3 R_*^{3/2} (GM_{\rm BH})^{-1/2}$.   Combining these yields
\begin{equation}
{\dot E}_{\rm peak} \propto  M_{\rm BH}^{3/2} \Phi^{-5} {\cal R}^{-5} R_*^{3/2} 
         \propto \Phi^{-5}(M_{*}^{8/3}/R_{*}^{5/2}) M_{\rm BH}^{-1/6}.
\end{equation}
The ratio between the predicted heating rate for our fiducial values and that found in the white dwarf simulation
is then $1.7 \times 10^{-4}  [(k/f)/0.08]^{-5/6} (M_*/M_{\odot})^{0.67} M_{\rm BH,6.5}^{-1/6}$ assuming $\xi = 0.2$.
Similarly, the timescale ratio is $8.3 \times 10^3  [(k/f)/0.08]^{1/2} M_{\rm BH,6.5}^{1/2} (M_*/M_{\odot})^{0.2}$.
Applying these factors to the heating rate calculated explicitly in \citet{Shiokawa+2015}, we find a heating rate as
a function of time shown in Figure~\ref{fig:heatingrate}.    For our fiducial parameters, it peaks at slightly less than
$10^{44}$~erg~s$^{-1}$, a factor  of $\sim 5$   below the order of magnitude estimate of eqn.~\ref{eq:L_peak}.
The characteristic time is $\simeq 21$~d, so the peak heating rate lasts for several months and then
trails off, roughly following the late-time decline of the mass-return rate.    These numbers are more illustrative than
general for two reasons: some details of TDE stream behavior scale with ${\cal R}$; and the simulation assumed a
Schwarzschild spacetime, while an orbital plane inclined relative to a Kerr spacetime could introduce further
complications \citep{Guillochon.Ramirez-Ruiz2015,Hayasaki15}. 

Whether this heat can be radiated quickly depends on the local optical depth.   Near the apocenter it is
\begin{equation}
\label{eq:tau}
\tau\sim \frac{\kappa M_*}{4\pi a_{\rm{min}}^2}\sim 500\,\left(\frac{k/f}{0.08}\right)^{-2/3}\cM^{1/3+2\xi}M_{\rm{BH,6.5}}^{-4/3},
\end{equation}
where we used the Thomson opacity $\kappa=0.34\,\rm{cm^2g^{-1}}$ and assumed that only half the star's mass is bound.   With such a large scattering optical depth, we assume the radiation is well thermalized and escapes with a roughly blackbody spectrum.   
If the vertical scale height $H \sim R$, the diffusion time $\tau H/c$ is
\begin{equation}
t_{\rm{diff}}\sim \frac{\tau a_{\rm{min}}}{c}\sim 6\times 10^6\,\left(\frac{k/f}{0.08}\right)^{-1/3}\cM^{2/3+\xi}M_{\rm{BH,6.5}}^{-2/3}\,\rm{s}.
\end{equation}
This time is, for our fiducial scenario, only a factor of three greater than the characteristic timescale of mass-return $t_0$, but its ratio to $t_0$ scales with black hole mass $\propto M_{BH}^{-7/6}$.    If, as is indicated by the simulation of \cite{Shiokawa+2015}, $H/R \simeq 0.5$, we expect the luminosity to track the heating rate for masses larger than $\sim 3 \times 10^6 M_{\odot}$, but be delayed with respect to the heating rate for smaller masses.   
In the quick diffusion time (large black hole mass) limit, the peak radiated bolometric luminosity is, up to the efficiency factor,   $\dot E_{\rm peak}$. 
This peak luminosity is always smaller than the Eddington luminosity $L_{Edd}$.   Although $\dot E_{\rm peak}/L_{Edd} \propto M_{BH}^{-7/6}$, the diffusion time grows slightly faster with decreasing mass.  For this reason, at lower masses $L_{\rm peak}/L_{Edd} \propto M_{BH}^{1/6}$, so that the maximum $L/L_{\rm Edd}$ is never more than a few tenths.   Instead, the relatively stronger heating rate when the black hole mass is smaller will likely lead to pushing matter outward (because the energy liberated is bounded by the binding energy, the majority of the mass cannot be expelled).

\begin{table*}
\centering
\bgroup
\def\arraystretch{1.5}
\begin{threeparttable}
\caption{\label{tab:predictions}Model predictions for TDEs with a rise time of a month}
\begin{tabular}{lcccccc}
\toprule
Stellar & $k/f$  & $M_{\rm{BH}}$      & $L_{\rm{peak}}$ & $T_{\rm{BB}}$ & $R_{\rm{BB}}$ & Line width \\
	structure	& & $(10^6M_{\odot})$  & $(10^{43}\rm{erg\,s^{-1}})$ & ($10^4$ K) & ($10^{15}$ cm) & ($\rm{km\,s^{-1}}$)  \\
		 \midrule
radiative & $0.02$ & 10 & 130 & 5.6 & 0.44 & 17000 \\
convective & $0.3$ & 1 & 20 & 4.8 & 0.23 & 7500 \\
fiducial   & $0.08$ & 3 & 50 & 5.1 & 0.31 & 11000 \\
\bottomrule
\end{tabular}
\begin{tablenotes}
\item The first prototype matches {predominantly radiative main sequence stars, those with ${\cal M}_* \simeq 1$.} The second prototype is for { mostly convective stars, which tend to be stars of either very low or very high mass.}  Our fiducial case, presented in the third line, is defined by the geometrical means of the extreme values for the $k/f$ factor and the black hole mass. {$L_{\rm{peak}}$ presented here and the corresponding temperature  (see Eqs. \ref{eq:L_peak} and \ref{eq:T}) assume 100\% efficiency of the shocks in converting gravitational energy to observed luminosity. The actual efficiency is lower, e.g. a comparison to \citealt{Shiokawa+2015} suggests luminosities lower by a factor of $\approx$5 and temperatures lower by a factor of $\approx$1.5 (see text).}
\end{tablenotes}
\end{threeparttable}
\egroup
\end{table*}

The pericenter shock also heats the gas, but the optical depth in this region is so much greater that its contribution to the luminosity is small.   Accumulation of only $10^{-4}$  of the  infalling matter near $R_T$ results in a  diffusion time from this region that is longer than the accretion time scale (for $\alpha = 0.1$).

The blackbody temperature of the apocenter radiation in the rapid diffusion limit is:
\begin{equation}\label{eq:T}
T \sim 5.1  \times 10^4~ {\rm K} 
\left(\frac{k/f}{0.08}\right)^{-3/8}\cM^{-\frac{1}{8}+\frac{9\xi}{8}}M_{\rm{BH,6.5}}^{-3/8}.
\end{equation}
The typical relative velocity between shocking streams at the apocenter region is the Keplerian velocity: 
\begin{equation}\label{eq:v}
v\approx\left(\frac{GM_{\rm{BH}}}{a_{\rm{min}}}\right)^{1/2}\approx 11000 \,\rm{km\,s^{-1}} \left(\frac{k/f}{0.08}\right)^{-1/6}\cM^{\xi/2-1/6}M_{\rm{BH,6.5}}^{1/6}.
\end{equation}

If the orbits were circular and the line emission confined to a narrow annulus, the line profile would have a pair of peaks separated by $2v\sin i$, for inclination angle $i$.    Elliptical motion can alter this separation by a factor of order unity and shift its center; a larger radial spread in the zone of emission can, to a degree, fill in the minimum in the line profile.   


Eqs. (\ref{eq:L_peak}) -  (\ref{eq:v}) agree reasonably  with the results of the simulation by \cite{Shiokawa+2015} when extrapolated from their simulated white dwarf scenario to our fiducial main sequence scenario.
According to the simulation by \cite{Shiokawa+2015}, although shock~1 initially dominates the dissipation rate, after a time $\sim t_0$, shocks~2 and 3 produce the most heat.  The rise of the outer shocks marks the beginning of efficient dissipation of orbital energy at the apocenter region, and it therefore corresponds to the rise of the signal that we model. Following the peak of the outer shocks' dissipation, the simulated energy dissipation rate decreases.  Although the heating rate falls below the peak rate by only a factor of 2 by the end of the simulation at $t \simeq 12t_0$, its decay is at least roughly consistent with  $t^{-5/3}$, which is what would be expected if late-time heating were primarily due to shocks acting on freshly-returned matter.

\begin{table*}
\centering
\bgroup
\def\arraystretch{1.5}
\begin{threeparttable}
\caption{\label{tab:observations}Observed properties of  optical TDE candidates}
\begin{tabular}{lccccc}
\toprule
Event    & $M_{\rm{BH}}$      & $L_{\rm{peak}}$ & $T_{\rm{BB}}$ & $R_{\rm{BB}}$ & Line width  \\
		 & $(10^6M_{\odot})$  & $(10^{43}\rm{erg\,s^{-1}})$ & ($10^4$ K) & ($10^{15}$ cm) & ($\rm{km\,s^{-1}}$)  \\
\midrule
PS1-10jh$^1$   & $4^{+4}_{-2}$  & $\gtrsim 22$  & $\gtrsim 3$  & $\sim 0.6$ & $9000\pm 700$  \\
PS1-11af$^2$   & $8\pm 2$  & $8.5\pm0.2$  & $1.90\pm0.07$  & $\sim 1.2$ &  \\
PTF09ge$^3$    & $5.65^{+3.02}_{-0.98}$  &  $85^{+50}_{-40}$ & $3.1\pm0.3$  & $1.14\pm0.2$ & $10070\pm 670$ \\
    &   &  $5.8^{+5.3\,\textbf{(a)}}_{-3.3}$ & $2.2\pm0.3$  & $0.59^{+0.16}_{-0.12}$ &   \\
SDSS TDE2$^4$  & $35.5^{+55.3}_{-25.8}$  & $\gg4.1^{\textbf{(b)}}$  & $1.82^{+0.07}_{-0.06}$  &  & $\sim$8000$^{\textbf{(c)}}$  \\
ASASSN14ae$^5$  & $2.45^{+1.55}_{-0.74}$ & $8.2\pm0.5$  &  $2.2\pm0.1$ & $0.7\pm0.03$ & 17000-8000$^{\textbf{(c)}}$  \\
PTF09axc$^3$  & $2.69^{+0.66}_{-0.64}$ &  $1.9^{+3.3\,\textbf{(d)}}_{-1.4}$ & $1.19^{+0.32}_{-0.17}$  & $1.14^{+0.41}_{-0.43}$ & $11890\pm220$  \\
PTF09djl$^3$  & $3.57^{+9.97}_{-2.96}$ & $12.7^{+23.1\,\textbf{(e)}}_{-10.4}$  & $2.7^{+0.7}_{-0.5}$  & $0.58^{+0.24}_{-0.21}$ & $6530\pm350$  \\

\bottomrule
\end{tabular}
\begin{tablenotes}
\item   (1)   \citealt{Gezari.et.al.2012} (2) \citealt{Chornock+2014} (3) \citealt{Arcavi+2014} (4) \citealt{vanVelzen2014} (5) \citealt{Holoien+2014}.
\item For the first four events  the authors reported a reasonable agreement between the observed light curve and a $\sim t^{-5/3}$ decline.  ASASSN-14ae does  not show such a decline while the last two are not clear. 
\item (a) PTF09ge: First line,  values we fitted from the peak band luminosities kindly supplied by Iair Arcavi; second line, published values for $L$, $T_{\rm{BB}}$ and $R_{\rm{BB}}$ at 19 days before peak \citep{Arcavi+2014}.
\item (b) The $L_{\rm{peak}}$ values presented are bolometric, except for SDSS TDE2, where it is only the g band. The bolometric peak for this event is not published, but is probably similar to or higher than PS1-10jh.
\item (c) For SDSS TDE2 and ASASSN-14ae, the values are from the discovery papers \citep{vanVelzen2014,Holoien+2014}. These should be compared with $3440\pm 1100$ and $3600\pm175$ (respectively), reported by \cite{Arcavi+2014}. For ASASSN-14ae, $17000\rightarrow8000$ means $17000\,\rm{km\,s^{-1}}$ at the luminosity peak and a decrease to $8000\,\rm{km\,s^{-1}}$ three months later.
\item (d) A value estimated using the published values of $T$ and $R$  \citep[determined by ][using a fit to the continuum at around the peak of the light curve]{Arcavi+2014}. Note that this is comparable to the r-band luminosity and thus should be taken as a lower limit. 
\item (e) A value estimated using the published values of $T$ and $R$ \citep[determined by ][using a fit to the continuum at around the peak of the light curve]{Arcavi+2014}.
\end{tablenotes}
\end{threeparttable}
\egroup
\end{table*}

Eqs. 
(\ref{Eq. amin}), (\ref{eq:t0}), (\ref{eq:L_peak}), (\ref{eq:T}) and (\ref{eq:v})
provide estimates for  the values of five observables (radius, time, $L_{peak}$, $T$ and  $v$) that characterize the emission due to the circularization process.  To illustrate the likely span of observable properties predicted by our model, we computed what we would expect from each of three sets of parameters, our fiducial set and two others, differing in stellar structure ($k/f$) and black hole mass $M_{\rm BH}$.  These sample predictions are shown in Table~\ref{tab:predictions}.

We emphasize that the signal discussed here is emitted regardless of whatever radiation is produced by other processes during the course of the TDE, e.g. the emission produced by the accretion process or by outflow-driven shocks.  For TDEs with $R_p/R_g$ at least a few tens, the principal remaining uncertainty is stellar structure; this uncertainty is reflected in the factor $k/f$.
On the other hand, there may be classes of TDEs to which our analysis does not apply.   For example, if $R_p/R_g \lesssim 10$, whether because $M_{BH}$ is especially large or $R_p$ is especially small, relativistic apsidal precession or, if the black hole rotates, Lense-Thirring precession may lead to strong shocks at radii nearer (logarithmically) to  $R_p$ than to $a_{\rm min}$.
In this case the circularization radius will be  smaller and  the circularization signal will be brighter and hotter. The circularization energy would then  be comparable to the accretion energy and this signal might blend with  the accretion signal.

\section{Comparison with TDE Candidates' Observations}
\label{sec:obs}

\cite{Arcavi+2014} describe seven rather similar optical TDE candidates.   All these candidates (by selection)
had  comparable peak luminosities and light curves with similar timescales.  
Table~\ref{tab:observations} presents a summary of these observations (including in addition PS1-11af, \citealt{Chornock+2014}, but excluding SDSS J0748, \citealt{Wang2011}, which has no reported $L_{\rm{peak}}$, $T$ and $R$). Presented in the table are, for each of these optical TDE candidates, the SMBH mass (as estimated by \citealt{Arcavi+2014} using the \citealt{Gadotti2009} and \citealt{HaringRix2004} bulge relations), the estimated peak  bolometric luminosity, the fitted blackbody temperature and radius, the width of observed HeII or H$\alpha$ emission lines, and whether the optical light curve is consistent with a  $t^{-5/3}$ decay trend.  For most of the optical candidates, the SMBH mass is a few times $10^6M_{\odot}$. 
For such SMBH masses disrupting a one solar mass main sequence star, our model predicts $L_{\rm{peak}} \sim 10^{44}\, \rm{erg\,s^{-1}}$, a blackbody temperature $\sim 4 \times10^4$~K, a blackbody radius (i.e., $a_{\rm min}$) $\sim 5 \times 10^{14}$ cm, and a line width $\sim 8000\,\rm{km\,s^{-1}}$.    These figures agree quite well with the measurements reported in Table~\ref{tab:observations}.

A unique feature of many optical TDE candidates is an observed temperature almost constant in time, in contrast with theoretical predictions of accretion theory. For example, the spectral shape of PS1-10jh \citep{Gezari.et.al.2012} at optical-NUV wavelengths is unchanged while the integrated flux falls by a full order of magnitude, and  the ASASSN-14ae \citep{Holoien+2014} fitted temperature remains $\sim 2\times 10^4$ K during a fall of one and a half orders of magnitude in integrated flux.   When the temperature is constant, the bolometric correction should be likewise constant. 
A decay in the luminosity combined with a constant $T$ then implies that the radius is decreasing with time. A possible explanation is a gradual shift inward of shock 2 as debris that already passed through pericenter gradually settles into less eccentric orbits.

One can easily see the general agreement between the observations reported in Table \ref{tab:observations} and the model predictions summarized in Eqs. 
(\ref{Eq. amin}), (\ref{eq:t0}), (\ref{eq:L_peak}), (\ref{eq:T}),  (\ref{eq:v})
and in Table \ref{tab:predictions}. 
{The values of the estimated  luminosities (and corresponding temperatures) are slightly larger than the 
observed  values. This is reasonable given that after all they are only and order of magnitude estimates which assume 100\% efficiency. Indeed a comparison with the scaled luminosities based on \cite{Shiokawa+2015} 
show indeed luminosities that are smaller by about a factor of 5. }
Although the observed values do not exactly match the values expected for our fiducial scenario,
we did not attempt to fit the model parameters to the observations.  
Both the model and the data (N.B. the absence of extinction corrections  and the discrepancies between different observations noted in Table~\ref{tab:observations}) are too crude for that. 
The rise time to peak, roughly reflecting $t_0$, is about a month for all events with a well observed rise phase. This may reflect a selection effect, as many of the events were found in data of surveys with cadences optimized for detection of supernovae.

 For any particular stellar structure, i.e., choice of $k/f$, any pair of the four observables $L_{\rm peak}$, $T$, $v$ and $t_0$ could in principle be used to solve for $M_{\rm BH}$ and $M_*$, thus over-constraining these values.  However, varying $M_*$ within a factor of a few, reasonable for main sequence stars, has only  a marginal effect on the observables, which depend very weakly on the mass of the tidally disrupted star.   Therefore, given a value of $k/f$, the observables' values are mostly determined by the SMBH mass.   For example, for our fiducial value of $k/f$, a rise time of  $t_0\sim1$ month implies $M_{\rm BH,6.5} \sim1$.   In addition, both larger $k/f$ and larger $M_{\rm BH}$ diminish the luminosity and temperature, but increase the line-width.   Consequently, to match a specific range of observables, a smaller $k/f$ (a star more of whose volume is radiative) demands a larger $M_{\rm BH}$, and vice versa.   However, the degree of sensitivity to these parameters varies; in particular, the line width and peak luminosity depend more strongly on $M_{\rm BH}$ than the temperature.   Thus, there are distinctions between the three cases listed in Table~\ref{tab:predictions}.

\section{Discussion}
\label{sec:conc} 

We suggest here that the observed optical light from TDEs arises from emission on radial scales $\sim a_{\rm min}$, and that the energy source is the orbital energy dissipated by shocks at that distance  from the central SMBH.   Although the energy dissipated in this process is a small fraction ($\sim 10^{-2}$) of the total energy available from accretion onto the black hole, 
it is actually of the right scale to power the observed signal.  The observed temperatures, emission radii and line widths are all in agreement with the expected characteristics of this emission.   The observed rate of decline is also crudely consistent with the decline in heating rate due to these shocks.   All this, plus the fact that radiation from the region of tidal stream shocks must occur in any event, are significant advantages.

At the same time, however, these successes point to a puzzle.   Our model says nothing about what happens to the $\sim 10^{53}$~erg one might expect to be released when the debris is accreted onto the black hole.   Where does this energy go?  In fact, this question is raised equally strongly in regard to models in which all the radiation stems directly from accretion.   Posed in that context, it becomes: why do we see only 1\% of the expected energy?    The only difference between the nature of this question as it applies to our model and as it applies to the more conventional models is that in our case the suppression factor could be anything smaller than $\sim 0.01$; in the conventional models, it must be consistently $\simeq 0.01$.

A number of speculative answers to this question exist.     One possibility is that the mass return rate can often be super-Eddington, suggesting strong photon trapping in the deepest part of the gravitational well, where most of the accretion energy is released \citep{Begelman1979, Abramowicz1988}.   This option raises several difficulties.   On the theoretical side, \cite{Shiokawa+2015} found that the peak accretion rate onto the black hole is reduced by an order of magnitude relative to conventional models, restricting the parameter space of super-Eddington accretion to lower-mass black holes.    In addition, recent simulational work on super-Eddington accretion has shown that magnetic buoyancy effects may provide an end-run around photon-trapping that permits the emerging luminosity to match the heating rate \citep{Jiang2014}.    On the observational side, if strong photon-trapping does occur, it would create a light curve with a lengthy period of nearly constant luminosity \citep{Krolik.Piran.2012}, and none of the events discussed here shows such behavior.   Indeed, the luminosity begins to decline within a month or two of the peak.   If photon-trapping explained the small radiated energy, this common behavior would require  fine-tuning, for all observed events would have to be just slightly super-Eddington. Note, however, that the peak mass-return rate in Eddington units is only $\simeq  40 M_*^{1+3\xi)/2} M_{\rm BH,6.5}^{-1/2}$ \citep{Shiokawa+2015}; there may be a population
of events with somewhat larger masses for which the mass-return rate is super-Eddington for only a short period of time. 
Another option is that when the mass accretion rate is super-Eddington, 99\% or more of the accretion power is put into kinetic energy of a low-mass outflow rather than photons \citep{Ayal+2000,Ohsuga+2005}.   In the super-Eddington accretion simulations of \cite{Sadowski+2014}, the effective radiative efficiency in rest-mass units at an accretion rate $100 \dot M_{Edd}$ (defined with respect to an assumed radiative efficiency of 0.057) falling into a non-spinning black hole was indeed depressed by about a factor of 100.   However, when they examined such a flow onto a black hole with spin parameter 0.9, they found a suppression factor of only about 20 and a substantial jet power.    Their simulation also raises another problem with stronger implications for this option: they predict a very strong anisotropy of the emitted photon luminosity.    One would then expect that surveys would strongly favor discovering systems in which the apparent radiative efficiency is much greater, and their spectra would be much hotter than those actually seen.

A third is that 99\% or more of the accretion flow is blown away before it ever comes near the black hole.    In one version of this suggestion, the wind prevents the accretion rate onto the black hole from significantly exceeding Eddington \citep{Poutanen+2007}.   In that case, it suffers from the same lightcurve problem as photon-trapping, for both amount to mechanisms that regulate the accretion onto the black hole to a steady near-Eddington level, for {\it any} value of the accretion rate greater than Eddington.   It has also been suggested that the mass-loss might be regulated by atomic line opacity \citep{LaorDavis2014,Miller2015}.   This version has the desirable property of predicting the luminosity to decrease as the accretion rate diminishes, but not as rapidly as the magnitude of the accretion rate in the outer disk falls.   On the other hand, although there are significant intrinsic uncertainties in this model's prediction of the luminosity, in rough terms it predicts a total radiated energy only a factor $\sim 10$ smaller than the conventional prediction because nearly all the accreting matter reaches radii $\sim 100R_g$

A fourth imaginable answer is that the additional energy is radiated away in the EUV/FUV, or non-thermally in a different channel that has not yet been observed.    
Such a situation would be expected if, when accretion onto the black hole finally begins, the light generated in the inner disk avoids reprocessing on its way out.
Another version of this explanation is that the actual extinction corrections are much larger than currently supposed.   Without further definition, it is hard to evaluate this class of solution.

Finally, we comment that the lengthened timescale for any particular combination of stellar mass and black hole mass predicted by \cite{Shiokawa+2015} bears on this question only in the sense that it diminishes the range of parameters for which super-Eddington accretion would be expected.    The total accretion energy predicted would be the same as any other model in which roughly half of the disrupted star is ultimately accreted onto the black hole.

We close with a few thoughts about observations that could provide tests for our model.
Our predicted peak luminosity depends very weakly on the black hole's mass, but the temperature decreases and the expected line-width increases when $M_{BH}$ increases.   Thus, a sample of events with well-determined black hole masses, peak luminosities, emission line widths, and characteristic temperatures would provide a strong statistical test.

We thank Iair Arcavi for his kind help in interpreting the data and Avishai Gal Yam, Eran Ofek, Re'em Sari and Rosemary Wyse for helpful discussion and remarks {and an anonymous referee for very instructive comments}. This work was partially supported by an ERC advanced grant ``GRBs",  by the I-CORE 
Program of the Planning and Budgeting Committee and The Israel Science Foundation grant No 1829/12 and by  ISA grant 3-10417  (TP). Additional support was received from
NSF grant AST-1028111 and NASA/ATP grant NNX14AB43G (JHK).

\bibliographystyle{apj}

\begin{thebibliography}{51}
\expandafter\ifx\csname natexlab\endcsname\relax\def\natexlab#1{#1}\fi

\bibitem[{{Abramowicz} {et~al.}(1988){Abramowicz}, {Czerny}, {Lasota}, \&
  {Szuszkiewicz}}]{Abramowicz1988}
{Abramowicz}, M.~A., {Czerny}, B., {Lasota}, J.~P., \& {Szuszkiewicz}, E. 1988,
  \apj, 332, 646

\bibitem[{{Arcavi} {et~al.}(2014){Arcavi}, {Gal-Yam}, {Sullivan}, {Pan},
  {Cenko}, {Horesh}, {Ofek}, {De Cia}, {Yan}, {Yang}, {Howell}, {Tal},
  {Kulkarni}, {Tendulkar}, {Tang}, {Xu}, {Sternberg}, {Cohen}, {Bloom},
  {Nugent}, {Kasliwal}, {Perley}, {Quimby}, {Miller}, {Theissen}, \&
  {Laher}}]{Arcavi+2014}
{Arcavi}, I., {et~al.} 2014, \apj, 793, 38

\bibitem[{{Ayal} {et~al.}(2000){Ayal}, {Livio}, \& {Piran}}]{Ayal+2000}
{Ayal}, S., {Livio}, M., \& {Piran}, T. 2000, \apj, 545, 772

\bibitem[{{Begelman}(1979)}]{Begelman1979}
{Begelman}, M.~C. 1979, \mnras, 187, 237

\bibitem[{{Bloom} {et~al.}(2011){Bloom}, {Giannios}, {Metzger}, {Cenko},
  {Perley}, {Butler}, {Tanvir}, {Levan}, {O'Brien}, {Strubbe}, {De Colle},
  {Ramirez-Ruiz}, {Lee}, {Nayakshin}, {Quataert}, {King}, {Cucchiara},
  {Guillochon}, {Bower}, {Fruchter}, {Morgan}, \& {van der Horst}}]{Bloom+2011}
{Bloom}, J.~S., {et~al.} 2011, Science, 333, 203

\bibitem[{{Bogdanovi{\'c}} {et~al.}(2014){Bogdanovi{\'c}}, {Cheng}, \&
  {Amaro-Seoane}}]{Bogdanovic+2014}
{Bogdanovi{\'c}}, T., {Cheng}, R.~M., \& {Amaro-Seoane}, P. 2014, \apj, 788, 99

\bibitem[{{Bonnerot} {et~al.}(2015){Bonnerot}, {Rossi}, {Lodato}, \&
  {Price}}]{Bonnerot+15}
{Bonnerot}, C., {Rossi}, E.~M., {Lodato}, G., \& {Price}, D.~J. 2015, ArXiv
  e-prints

\bibitem[{{Chornock} {et~al.}(2014){Chornock}, {Berger}, {Gezari}, {Zauderer},
  {Rest}, {Chomiuk}, {Kamble}, {Soderberg}, {Czekala}, {Dittmann}, {Drout},
  {Foley}, {Fong}, {Huber}, {Kirshner}, {Lawrence}, {Lunnan}, {Marion},
  {Narayan}, {Riess}, {Roth}, {Sanders}, {Scolnic}, {Smartt}, {Smith},
  {Stubbs}, {Tonry}, {Burgett}, {Chambers}, {Flewelling}, {Hodapp}, {Kaiser},
  {Magnier}, {Martin}, {Neill}, {Price}, \& {Wainscoat}}]{Chornock+2014}
{Chornock}, R., {et~al.} 2014, \apj, 780, 44

\bibitem[{{Donley} {et~al.}(2002){Donley}, {Brandt}, {Eracleous}, \&
  {Boller}}]{Donley+2002}
{Donley}, J.~L., {Brandt}, W.~N., {Eracleous}, M., \& {Boller}, T. 2002, \aj,
  124, 1308

\bibitem[{{Dotan} \& {Shaviv}(2011)}]{Dotan.Shaviv2011}
{Dotan}, C., \& {Shaviv}, N.~J. 2011, \mnras, 413, 1623

\bibitem[{{Evans} \& {Kochanek}(1989)}]{Evans.Kochanek.1989}
{Evans}, C.~R., \& {Kochanek}, C.~S. 1989, \apjl, 346, L13

\bibitem[{{Frank} \& {Rees}(1976)}]{FrankRees1976}
{Frank}, J., \& {Rees}, M.~J. 1976, \mnras, 176, 633

\bibitem[{{Gadotti}(2009)}]{Gadotti2009}
{Gadotti}, D.~A. 2009, \mnras, 393, 1531

\bibitem[{{Gezari} {et~al.}(2009){Gezari}, {Heckman}, {Cenko}, {Eracleous},
  {Forster}, {Gon{\c c}alves}, {Martin}, {Morrissey}, {Neff}, {Seibert},
  {Schiminovich}, \& {Wyder}}]{Gezari+2009}
{Gezari}, S., {et~al.} 2009, \apj, 698, 1367

\bibitem[{{Gezari} {et~al.}(2012){Gezari}, {Chornock}, {Rest}, {Huber},
  {Forster}, {Berger}, {Challis}, {Neill}, {Martin}, {Heckman}, {Lawrence},
  {Norman}, {Narayan}, {Foley}, {Marion}, {Scolnic}, {Chomiuk}, {Soderberg},
  {Smith}, {Kirshner}, {Riess}, {Smartt}, {Stubbs}, {Tonry}, {Wood-Vasey},
  {Burgett}, {Chambers}, {Grav}, {Heasley}, {Kaiser}, {Kudritzki}, {Magnier},
  {Morgan}, \& {Price}}]{Gezari.et.al.2012}
---. 2012, \nat, 485, 217

\bibitem[{{Gonz{\'a}lez Delgado} {et~al.}(2014){Gonz{\'a}lez Delgado},
  {Garc{\'{\i}}a-Benito}, {P{\'e}rez}, {Cid Fernandes}, {de Amorim},
  {Cortijo-Ferrero}, {Lacerda}, {L{\'o}pez Fern{\'a}ndez}, {S{\'a}nchez}, {Vale
  Asari}, \& {CALIFA collaboration}}]{Gonzalez+2014}
{Gonz{\'a}lez Delgado}, R.~M., {et~al.} 2014, ArXiv e-prints

\bibitem[{{Guillochon} {et~al.}(2014){Guillochon}, {Manukian}, \&
  {Ramirez-Ruiz}}]{Guillochon.Manukian.Ramirez-Ruiz2014}
{Guillochon}, J., {Manukian}, H., \& {Ramirez-Ruiz}, E. 2014, \apj, 783, 23

\bibitem[{{Guillochon} \& {Ramirez-Ruiz}(2013)}]{Guillochon.Ramirez-Ruiz.2013}
{Guillochon}, J., \& {Ramirez-Ruiz}, E. 2013, \apj, 767, 25

\bibitem[{{Guillochon} \& {Ramirez-Ruiz}(2015)}]{Guillochon.Ramirez-Ruiz2015}
---. 2015, ArXiv e-prints

\bibitem[{{H{\"a}ring} \& {Rix}(2004)}]{HaringRix2004}
{H{\"a}ring}, N., \& {Rix}, H.-W. 2004, \apjl, 604, L89

\bibitem[{{Hayasaki} {et~al.}(2015){Hayasaki}, {Stone}, \& {Loeb}}]{Hayasaki15}
{Hayasaki}, K., {Stone}, N.~C., \& {Loeb}, A. 2015, ArXiv e-prints

\bibitem[{{Holoien} {et~al.}(2014){Holoien}, {Prieto}, {Bersier}, {Kochanek},
  {Stanek}, {Shappee}, {Grupe}, {Basu}, {Beacom}, {Brimacombe}, {Brown},
  {Davis}, {Jencson}, {Pojmanski}, \& {Szczygie{\l}}}]{Holoien+2014}
{Holoien}, T.~W.-S., {et~al.} 2014, \mnras, 445, 3263

\bibitem[{{Jiang} {et~al.}(2014){Jiang}, {Stone}, \& {Davis}}]{Jiang2014}
{Jiang}, Y.-F., {Stone}, J.~M., \& {Davis}, S.~W. 2014, \apj, 796, 106

\bibitem[{{Kesden}(2012)}]{Kesden2012}
{Kesden}, M. 2012, \prd, 85, 024037

\bibitem[{{Kippenhahn} \& {Weigert}(1994)}]{Kippenhahn1994}
{Kippenhahn}, R., \& {Weigert}, A. 1994, {Stellar Structure and Evolution}
  (Springer-Verlag Berlin Heidelberg New York.)

\bibitem[{{Kochanek}(1994)}]{Kochanek1994}
{Kochanek}, C.~S. 1994, \apj, 422, 508

\bibitem[{{Komossa} {et~al.}(2004){Komossa}, {Halpern}, {Schartel}, {Hasinger},
  {Santos-Lleo}, \& {Predehl}}]{Komossa+2004}
{Komossa}, S., {Halpern}, J., {Schartel}, N., {Hasinger}, G., {Santos-Lleo},
  M., \& {Predehl}, P. 2004, \apjl, 603, L17

\bibitem[{{Krolik} \& {Piran}(2012)}]{Krolik.Piran.2012}
{Krolik}, J.~H., \& {Piran}, T. 2012, \apj, 749, 92

\bibitem[{{Laor} \& {Davis}(2014)}]{LaorDavis2014}
{Laor}, A., \& {Davis}, S.~W. 2014, \mnras, 438, 3024

\bibitem[{{Lodato}(2012)}]{Lodato12}
{Lodato}, G. 2012, in European Physical Journal Web of Conferences, Vol.~39,
  European Physical Journal Web of Conferences, 1001

\bibitem[{{Lodato} {et~al.}(2009){Lodato}, {King}, \&
  {Pringle}}]{Lodato.et.al.2009}
{Lodato}, G., {King}, A.~R., \& {Pringle}, J.~E. 2009, \mnras, 392, 332

\bibitem[{{Lodato} \& {Rossi}(2011)}]{Lodato2011}
{Lodato}, G., \& {Rossi}, E.~M. 2011, \mnras, 410, 359

\bibitem[{{Loeb} \& {Ulmer}(1997)}]{Loeb.Ulmer.1997}
{Loeb}, A., \& {Ulmer}, A. 1997, \apj, 489, 573

\bibitem[{{Magorrian} \& {Tremaine}(1999)}]{MagorrianTremaine99}
{Magorrian}, J., \& {Tremaine}, S. 1999, \mnras, 309, 447

\bibitem[{{Miller}(2015)}]{Miller2015}
{Miller}, M.~C. 2015, ArXiv e-prints

\bibitem[{{Ohsuga} {et~al.}(2005){Ohsuga}, {Mori}, {Nakamoto}, \&
  {Mineshige}}]{Ohsuga+2005}
{Ohsuga}, K., {Mori}, M., {Nakamoto}, T., \& {Mineshige}, S. 2005, \apj, 628,
  368

\bibitem[{{Phinney}(1989)}]{Phinney.1989}
{Phinney}, E.~S. 1989, in IAU Symposium, Vol. 136, The Center of the Galaxy,
  ed. M.~{Morris}, 543

\bibitem[{{Poutanen} {et~al.}(2007){Poutanen}, {Lipunova}, {Fabrika},
  {Butkevich}, \& {Abolmasov}}]{Poutanen+2007}
{Poutanen}, J., {Lipunova}, G., {Fabrika}, S., {Butkevich}, A.~G., \&
  {Abolmasov}, P. 2007, \mnras, 377, 1187

\bibitem[{{Rees}(1988)}]{Rees.1988}
{Rees}, M.~J. 1988, \nat, 333, 523

\bibitem[{{Rojas-Arriagada} {et~al.}(2014){Rojas-Arriagada}, {Recio-Blanco},
  {Hill}, {de Laverny}, {Schultheis}, {Babusiaux}, {Zoccali}, {Minniti},
  {Gonzalez}, {Feltzing}, {Gilmore}, {Randich}, {Vallenari}, {Alfaro},
  {Bensby}, {Bragaglia}, {Flaccomio}, {Lanzafame}, {Pancino}, {Smiljanic},
  {Bergemann}, {Costado}, {Damiani}, {Hourihane}, {Jofr{\'e}}, {Lardo},
  {Magrini}, {Maiorca}, {Morbidelli}, {Sbordone}, {Worley}, {Zaggia}, \&
  {Wyse}}]{Rojas+2014}
{Rojas-Arriagada}, A., {et~al.} 2014, \aap, 569, A103

\bibitem[{{Rosswog} {et~al.}(2009){Rosswog}, {Ramirez-Ruiz}, \&
  {Hix}}]{Rosswog.et.al.2009}
{Rosswog}, S., {Ramirez-Ruiz}, E., \& {Hix}, W.~R. 2009, \apj, 695, 404

\bibitem[{{Sarzi} {et~al.}(2005){Sarzi}, {Rix}, {Shields}, {Ho}, {Barth},
  {Rudnick}, {Filippenko}, \& {Sargent}}]{Sarzi+2005}
{Sarzi}, M., {Rix}, H.-W., {Shields}, J.~C., {Ho}, L.~C., {Barth}, A.~J.,
  {Rudnick}, G., {Filippenko}, A.~V., \& {Sargent}, W.~L.~W. 2005, \apj, 628,
  169

\bibitem[{{S{\c a}dowski} {et~al.}(2014){S{\c a}dowski}, {Narayan}, {McKinney},
  \& {Tchekhovskoy}}]{Sadowski+2014}
{S{\c a}dowski}, A., {Narayan}, R., {McKinney}, J.~C., \& {Tchekhovskoy}, A.
  2014, \mnras, 439, 503

\bibitem[{{Shiokawa} {et~al.}(2015){Shiokawa}, {Krolik}, {Cheng}, {Piran}, \&
  {Noble}}]{Shiokawa+2015}
{Shiokawa}, H., {Krolik}, J.~H., {Cheng}, R.~M., {Piran}, T., \& {Noble}, S.~C.
  2015, ArXiv e-prints

\bibitem[{{Stone} {et~al.}(2013){Stone}, {Sari}, \&
  {Loeb}}]{Stone.Sari.Loeb2013}
{Stone}, N., {Sari}, R., \& {Loeb}, A. 2013, \mnras, 435, 1809

\bibitem[{{Stone} \& {Metzger}(2014)}]{Stone2014}
{Stone}, N.~C., \& {Metzger}, B.~D. 2014, ArXiv e-prints

\bibitem[{{Strubbe} \& {Quataert}(2009)}]{Strubbe.Quataert.2009}
{Strubbe}, L.~E., \& {Quataert}, E. 2009, \mnras, 400, 2070

\bibitem[{{Ulmer}(1999)}]{Ulmer99}
{Ulmer}, A. 1999, \apj, 514, 180

\bibitem[{{van Velzen} \& {Farrar}(2014)}]{vanVelzen2014}
{van Velzen}, S., \& {Farrar}, G.~R. 2014, ArXiv e-prints

\bibitem[{{Wang} \& {Merritt}(2004)}]{Wang.Merritt.2004}
{Wang}, J., \& {Merritt}, D. 2004, \apj, 600, 149

\bibitem[{{Wang} {et~al.}(2011){Wang}, {Zhou}, {Wang}, {Lu}, \&
  {Xu}}]{Wang2011}
{Wang}, T.-G., {Zhou}, H.-Y., {Wang}, L.-F., {Lu}, H.-L., \& {Xu}, D. 2011,
  \apj, 740, 85

\end{thebibliography}

\end{document}